\documentclass[twocolumn,preprintnumbers, amssymb,amsmath,aps,floatfix,prl,nofootinbib,superscriptaddress,showpacs]{revtex4-1}

\usepackage{epsfig}
\usepackage{bm}
\usepackage{amssymb}
\usepackage{amsmath}
\usepackage{color}
\usepackage{subfigure}
\usepackage[colorlinks,
            linkcolor=blue,
            anchorcolor=black,
            citecolor=blue
            ]{hyperref}

\newcommand{\beq}{\begin{eqnarray}}
\newcommand{\eeq}{\end{eqnarray}}

\begin{document}
\title{Gluonic Probe for the Short Range Correlation in Nucleus}

\author{Ji Xu}
\affiliation{Department of Physics and Astronomy, Shanghai Jiao Tong University, Shanghai, China}
\affiliation{Nuclear Science Division, Lawrence Berkeley National
Laboratory, Berkeley, CA 94720, USA}

\author{Feng Yuan}
\affiliation{Nuclear Science Division, Lawrence Berkeley National
Laboratory, Berkeley, CA 94720, USA}

\begin{abstract}
We investigate the gluonic probe to the nucleon-nucleon short range correlation (SRC) in nucleus through heavy flavor production in deep inelastic scattering (DIS). The relevant EMC effects of $F_2^{c\bar c}$ structure function will provide a universality test of the SRCs which have been extensively studied in the quark-channel. These SRCs can also be studied through the sub-threshold production of heavy flavor in $eA$ collisions at the intermediate energy electron-ion collider, including open Charm and $J/\psi$ ($\Upsilon$) production.
\end{abstract}
\maketitle

\section{Introduction}

There have been renaissance interests to investigate the short-range nucleon nucleon correlation (SRC) of nucleus in recent years~\cite{Frankfurt:2008zv,Arrington:2011xs,Hen:2013oha,Hen:2016kwk,Fomin:2017ydn,Cloet:2019mql}. The SRC describes the probability that the two nucleons are close in coordinate space, as a result of nontrivial nucleon-nucleon interactions in nucleus. Especially, the experiment efforts made at JLab have stimulated much of the research activities in the field. The connection between the SRC and the well-known EMC effects in nuclear structure function measurements have been extensively investigated with compelling evidences~\cite{Egiyan:2005hs,Seely:2009gt,Weinstein:2010rt,Fomin:2011ng,Hen:2012fm,Arrington:2012ax,Hen:2014nza,Cohen:2018gzh,Duer:2018sby,Duer:2018sxh,Schmookler:2019nvf}. Since the EMC effects concerns the parton distribution functions through short distance physics while the SRC represents the wave function information of nucleons inside the nucleus, this relation, if finally established rigorously, shall provide a unique method to study nuclear structure physics and help to answer the long standing quest to explore nuclear structure through the first principle of strong interaction QCD theory, see, for example, some of recent developments in Refs.~\cite{Chen:2016bde,Lynn:2019vwp,Lynn:2019rdt}. 

One of the key aspects of the connection between the SRC and EMC effects is the universality, where the partonic structure from the correlated nucleon-nucleon in the nucleus contribute to both phenomena~\cite{Schmookler:2019nvf,Chen:2016bde,Lynn:2019vwp,Lynn:2019rdt,Frankfurt:1993sp,Piasetzky:2006ai,Segarra:2019gbp}. Therefore, to fully establish the physics case of the connection, we need to build a rigorous test on the universality. In this paper, we propose a number of novel observables for this purpose, focusing on the gluonic channel. 
We will tackle this issue from two fronts, both through heavy flavor production. First, we will discuss the nuclear modification of the gluon distribution function in the EMC region. Heavy flavor production in DIS is sensitive to the gluon distribution, and the nuclear modification will provide crucial information on the gluon distribution in nucleus. This can be studied by measuring the charm-structure function: $F_2^{c\bar c}$ and $F_L^{c\bar c}$. We will demonstrate the sensitivities for an intermediate energy EIC (IEEIC), where the main kinematic focuses are in the EMC region, i.e., $0.3<x<0.7$~\cite{NuXu}. Meanwhile, the nuclear modification for the gluon distribution has been a subject for decades' study, and has been widely applied in high energy processes~\cite{Eskola:2016oht,Kovarik:2015cma,Khanpour:2016pph,deFlorian:2011fp,AbdulKhalek:2019mzd,Walt:2019slu}. The measurements at the future IEEIC and the EIC~\cite{Aschenauer:2017oxs} will provide an important contribution to constrain the gluon distribution in nucleus.

Second, we propose to study the SRC through the sub-threshold heavy flavor production. This can be done by measuring either open Charm or Charmonium production in $eA$ collisions below the $ep$ threshold. This is similar to the structure function measurements beyond $x_B\sim 1$. 

Together with the quark sector study, the gluonic probe provides an important test of the universality of the SRC. In addition, because of the isospin symmetry, the nuclear modification can be derived directly without considering the isospin dependence as that for the quark sector. This shall provide a crucial way to disentangle different methods, for example, to extract the nuclear EMC effects~\cite{Arrington:2019wky,Hen:2019jzn}. 

The rest of this paper is organized as follows. We first study the charm structure function and the nuclear effects in the EMC region. We apply the EPPS16 parameterization for the gold nucleus and demonstrate the sensitivity to study the EMC effects in the gluonic channel. To illustrate the universality feature, we show the nuclear modification on the Charm structure function measurements for different nuclei based on the universal SRC contributions. The associated nuclear modifications become identical when divided by the SRC factor, similar to that have been shown in the structure function measurements~\cite{Schmookler:2019nvf,Segarra:2019gbp}. We then study the sub-threshold production of heavy flavor in $eA$ collisions, taking the example of $J/\psi$ production. We follow recent developments in both theory and experiment of near-threshold $J/\psi$ production and make a model to estimate the sub-threshold production of $J/\psi$ production in $eA$ collisions. The ratio of the nuclear cross sections in this region can be used to study the universal SRC. This can be straightforwardly extended to $\Upsilon$s and open Charm and Bottom production. This composites into a class of universality test for the SRC. Finally, we summarize our paper and comment on the future developments.

\section{EMC Effects in Charm Structure Function $F_2^{c\bar c}$.}

According to the universality of the SRC, we can parameterize the nuclear gluon distribution in the EMC region as that for the structure function~\cite{Frankfurt:1993sp,Chen:2016bde,Segarra:2019gbp},
\begin{equation}
    g_A(x,Q^2)=Ag_p(x,Q^2)+2n_{src}^A \delta \tilde{g}(x,Q^2) \ , 
\end{equation}
where $n_{src}$ represents number of $np$ pair in nucleus $A$. In the above equation, we have applied the charge symmetry between the gluon distributions in the proton and neutron, i.e., $g_p(x,Q^2)=g_n(x,Q^2)$ (see, for example, experimental measurement of this in Ref.~\cite{Zhu:2007aa}), and $\delta \tilde{g}(x,Q^2)$ represents the difference between the gluon distribution in the proton-neutron pair in the SRC and the free nucleon.  In practive, we can also parameterize nuclear modification of the gluon distribution in term of the gluon distirbution in a free nucleon,
\begin{equation}
R_g^A=\frac{g_A(x,Q^2)}{Ag_p(x,Q^2)} \ ,
\end{equation}
and from this, we have
\begin{equation}
\frac{\delta\tilde{g}(x,Q^2)}{g_p(x,Q^2)}=\frac{R_g^A(x,Q^2)-1}{2n_{src}^A/A} \ .\label{unigluon}
\end{equation}
The universality of the SRC contribution leads to a universal function in the left hand side of the above equation, meaning that the right hand side does not depend on the type of nucleus. Therefore, we can parameterize the gluon distribution in nucleus $A$ in terms of that in $B$, 
\begin{eqnarray}\label{correlation_between_RgAiandRgAu}
  R_g^{A}(x,Q^2)=\frac{a_2^A}{a_2^B}\left[ R_g^{B}(x,Q^2)-1 \right]+1 \,,
\end{eqnarray}
where $a_2$ is defined as $a_2^A=(n_{src}^A/A)/(n_{src}^d/2)$ and represents the SRC ratio of nucleus $A$ respect to that of deuteron. In the literature $a_2^A$ is also labeled as $a_2(A/d)$. This ratio can also be measured through the nuclear structure function beyond $x_B\sim 1$ region~\cite{Weinstein:2010rt,Fomin:2011ng,Hen:2012fm,Arrington:2012ax}. 

Similar to Ref.~\cite{Aschenauer:2017oxs}, we compute the reduced cross section for Charm structure functions,
\begin{eqnarray}
  \sigma_{red}^{c\bar c}(x_B,Q^2)&=&\left(\frac{d\sigma^{c\bar c}}{dx_BdQ^2}\right)\frac{x_BQ^4}{2\pi \alpha^2[1+(1-y)^2]}\\
  &=&F_2^{c\bar c}(x_B,Q^2)-\frac{y^2}{1+(1-y)^2}F_L^{c\bar c}(x_B,Q^2) \ ,\nonumber
\end{eqnarray}
where $F_2$ and $F_L$ are Charm structure functions depending on the gluon distribution functions. In our numeric calculations, we apply the leading order perturbative results to illustrate the main physics sensitivities at an IEEIC. For the typical kinematics of $Q^2\sim 10\rm GeV^2$ and $x\sim 0.1$ we find that the reduced cross section for $ep$ is about $10^{-3}$ which corresponds to a total cross section of $10^4\rm fb/GeV^2$ at the kinematics of an EIC in China (EicC)~\cite{NuXu}. This shows that the EicC has a great potential to explore the nuclear EMC effects for the gluon distributions. 

\begin{figure}[h]
\centering
\includegraphics[width=0.8\columnwidth]{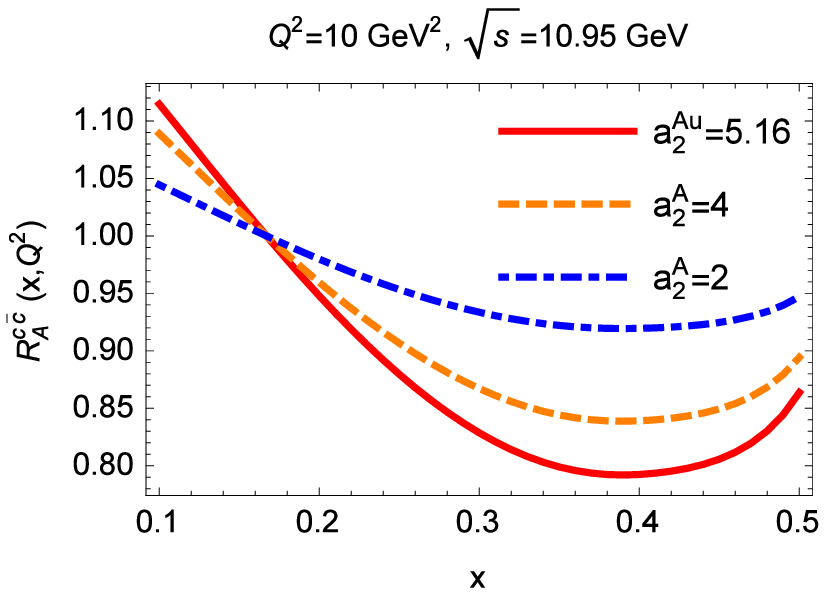}
\includegraphics[width=0.8\columnwidth]{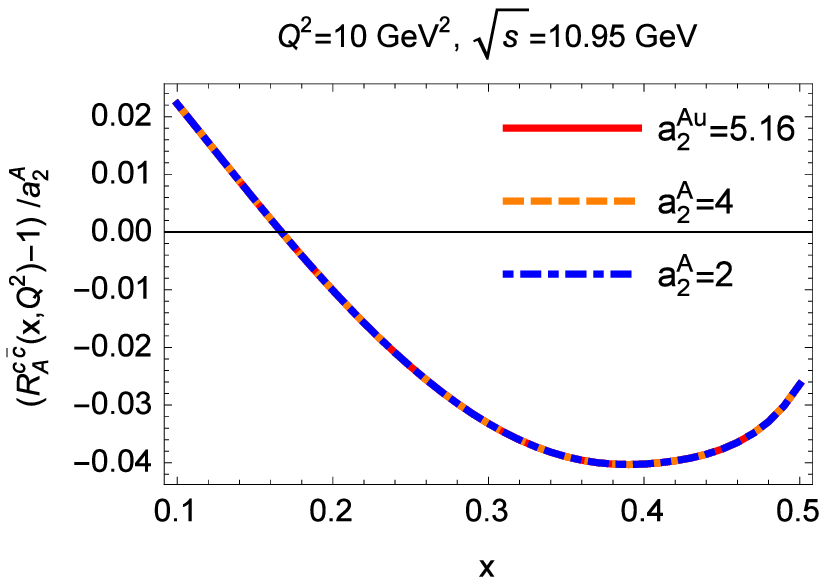}
\centering
\caption{Universality of EMC effects in the Charm structure function measurements at future intermediate energy EIC: (upper) $R_A^{c\bar c}(x,Q^2)$ defined in  Eq.(\ref{dif_of_RAcbarc}) and (lower) $\left[R_A^{c\bar c}-1\right]/a_2^A$ for different $a_2^A$: the solid red, orange and blue lines correspond to $a_2^{Au}=5.16$, $a_2^A=4$ and $a_2^A=2$ respectively.}
\label{ShownofRgAi}
\end{figure}

From the cross sections for $ep$ and $eA$ collisions, we can evaluate the nuclear modification,
\begin{equation}
R_A^{c\bar c}(x,Q^2)=\frac{\sigma_{A/red}^{c\bar c}(x,Q^2)}{A\sigma_{red}^{c\bar c}(x,Q^2)} \ .\label{dif_of_RAcbarc}
\end{equation}
As an example, in Fig.~\ref{ShownofRgAi} we show the above ratio for the gold nucleus, where we follow the EPPS16 parameterization~\cite{Eskola:2016oht} to illustrate the sensitivity. Notice that we do not have strong constraints on the gluon EMC effects from previous experiments. Therefore, the parameterizaztion from EPPS16 is only for illustration purpose. From this plot, we find that the nuclear factor 1.1 to 0.8 in the range of $x_B=0.1-0.4$. It is important to note that the actual momentum fraction carried by the gluon is larger than $x_B$. This can be estiamted as $x\approx x_B\tau$ with $\tau=(1+4m_c^2/Q^2)$. Therefore, the real EMC region is from 0.15 to 0.35 in this plot.

To investigate the universality of Eq.~(\ref{unigluon}), we need to explore different nuclei. For this, we parameterize the gluon distribution functions in two nuclei with different SRC factors: $a_2^A=4$ and $2$, and compute the $R_A^{c\bar c}$ ratios accordingly. We have plotted these ratios in Fig.~\ref{ShownofRgAi} as well. Future studies of these nuclear modification factors in the kinematics of Fig.~\ref{ShownofRgAi} will provide a universality test of the SRC.

This universality is better illustrated if we normalize the nuclear modification factors by the respective SRC factors~\cite{Schmookler:2019nvf,Segarra:2019gbp}. For that, we plot $\left[R_A^{c\bar c}-1\right]/a_2^A$ in the lower panel of Fig.~\ref{ShownofRgAi}. We clearly see that the three different curves in the upper panel shrink into one single line in the lower panel. This is mainly because the reduced cross sections are directly proportional to the associated gluon distributions: $R_A^{c\bar c}(x,Q^2)\propto R_g^A(\tau x,Q^2)$. According to the universality Eq.~(\ref{correlation_between_RgAiandRgAu}), we should have a universal function of $\left[R_A^{c\bar c}-1\right]/a_2^A$. That is a remarkable prediction from the universality of the SRC contributions. 

We would like to emphasize that the above studies are based on a leading order picture. Although we expect the above simple and intuitive picture will not change dramatically with higher order corrections, it will be interesting to see how the next-to-leading order computations will affect the universal behavior of the lower panel in Fig.~\ref{ShownofRgAi}. In terms of constraining the gluon distribution in nucleus, the measurements from the future IEEIC shall provide complementary information to that from an EIC~\cite{Aschenauer:2017oxs}.

\section{Sub-threshold Heavy Flavor Production at JLab and EIC}

The near threshold $J/\psi$ production in $\gamma p$ collisions has gained tremendous attentions in the last few years. One focus in these studies~\cite{Peskin:1979va,Bhanot:1979vb,Luke:1992tm,Kharzeev:1998bz,Kharzeev:1999jt,Brodsky:2000zc,Frankfurt:2002ka,Gryniuk:2016mpk,Hatta:2018ina,Hatta:2019lxo} is that it may provide important information on the proton mass decomposition~\cite{Ji:1994av}. At an EIC, we can also study the nuclear modification of the cross section near the threshold. It is more interesting to investigate the so-called sub-threshold $J/\psi$ production in $\gamma A$ collisions. The sub-threshold is a kinematic region that the individual $\gamma p$ collisions can not produce $J/\psi$, however, it is kinematically allowed to produce $J/\psi$ in $\gamma A$ collision. An important contribution is the SRC contributions, which is very similar to the case of the nuclear structure function beyond $x_B\sim 1$. 

Similar to the previous case, we have the following expression for $J/\psi$ production in $\gamma A$ process, 
\begin{eqnarray}
    &&\sigma_{\gamma A\to J/\psi}(W_{\gamma p})=A\sigma_{\gamma p\to J/\psi}(W_{\gamma p})\nonumber\\
    &&~~~+n_{src}^A(\sigma_{\gamma (pn)\to J/\psi}(W_{\gamma p})-2\sigma_{\gamma p\to J/\psi}(W_{\gamma p})) \ ,\label{xspsi}
\end{eqnarray}
where $W_{\gamma p}=\sqrt{s_{\gamma p}}$ represents the center of mass energy of $\gamma p$ and we have limited to two-body SRC. Again, we have applied the isospin symmetry to simplify the cross section calculations. We have also neglected the nuclear absorption correction for $J/\psi$ production (order of few percents)~\cite{Anderson:1976hi}, which could lead to suppression for both terms in the above equation. 

If the $\gamma p$ center of mass energy is below the $J/\psi$-threshold, the only contribution comes from the $\gamma (pn)$-term in the above,
\begin{equation}
    \sigma_{\gamma A\to J/\psi}(W_{\gamma p}<M_{pJ/\psi})=n_{src}^A\sigma_{\gamma (pn)\to J/\psi}(W_{\gamma p})  \ ,
\end{equation}
where have used $M_{pJ/\psi}=M_p+M_{J/\psi}$ to denote the total mass of $J/\psi$ and proton.
Therefore, in the sub-threshold region, the cross section ratios between different nuclei will not depend on the collision energy,
\begin{equation}
    \frac{\tilde\sigma_{\gamma A\to J/\psi}}{\tilde\sigma_{\gamma B\to J/\psi}}\big|_{W_{\gamma p}<M_{pJ/\psi}}=\frac{n_{src}^A}{n_{src}^B} \ ,
\end{equation}
where we have included a nuclear absorption corrections to $J/\psi$ production cross section and define $\tilde\sigma_{\gamma A\to J/\psi}=\sigma_{\gamma A\to J/\psi}/R_{abs.}^{A}$ for both $A$ and $B$ targets. In $\gamma A$ collisions, the short-distance produced $J/\psi$ could interact with other nucleons inside the nucleus, and results into a suppression (order of a few percents)~\cite{Anderson:1976hi}. 
In particular, if we take $B$-target as the deuterium, the above ratio will be the same as the structure function ratio beyond $x_B\sim 1$,
\begin{eqnarray}
    \frac{\tilde\sigma_{\gamma A\to J/\psi}}{\sigma_{\gamma d\to J/\psi}}\big|_{W_{\gamma p}<M_{pJ/\psi}}&=&\frac{n_{src}^A}{n_{src}^d}\nonumber\\
    &=&\frac{F_2^A(x_B)}{F_2^d(x_B)}|_{1.5<x_B<2.0} \ .
\end{eqnarray}
The latter ratios have been measured at previous DIS experiments with nuclear targets. The future measurements of the sub-threshold $J/\psi$ production will prove the universality of the SRC in these nucleus targets.

To estimate the cross sections for the sub-threshold production, we introduce the so-called energy fraction parameter $\chi_\gamma=\frac{M_{J/\psi}^2}{2E_\gamma M_p}+\frac{M_{J/\psi}}{E_\gamma}$~\cite{Brodsky:2000zc}, where $E_\gamma$ is the photon energy in the nucleus rest frame. The threshold limit of $\gamma p\to J/\psi+p$ corresponds to $\chi_\gamma\to 1$ limit. Therefore, we can apply a simple power behavior of $(1-\chi_\gamma)$ for the near threshold $J/\psi$ production, 
\begin{equation}
    \sigma_{\gamma p\to J/\psi}(W_{\gamma p})= \sigma_0^{\gamma p}
    (1-\chi_\gamma)^\beta \ , 
\end{equation}
where $\sigma_0=11.3~\rm nb$ and $\beta=1.3$ to represent the threshold behavior~\footnote{This also corresponds to the power behavior between the two-gluon exchange and three-gluon exchange models in Ref.\cite{Brodsky:2000zc}.}. This gives a very good description of previous experimental data~\cite{Gittelman:1975ix,Camerini:1975cy,Ali:2019lzf} in the near threshold region, and especially the very recent data from GlueX collaboration at JLab~\cite{Ali:2019lzf}.
It is also consistent, at least near the threshold region, with the calculation based on the AdS/CFT correspondence~\cite{Hatta:2018ina,Hatta:2019lxo}. Going to the case of $\gamma(pn)\to J/\psi$, we argue the basic formula remain and we may just only need to change the $\chi_\gamma$,
\begin{equation}
\chi_{\gamma}\to \tilde \chi_\gamma=\frac{M_{J/\psi}^2}{2E_\gamma 2M_p}+\frac{M_{J/\psi}}{E_\gamma} \ ,
\end{equation}
where we have used two-nucleon mass to represent the threshold kinematics. Therefore, we assume the SRC cross section can be written in the same form,
\begin{equation}
    \sigma_{\gamma (pn)\to J/\psi}(W_{\gamma p})= \sigma_0^{\gamma (pn)}
    (1-\tilde \chi_\gamma)^{\beta_2} \ .
\end{equation}
Because at higher energy $(1-\tilde \chi_\gamma)$ becomes $(1-\chi_\gamma)$, we further assume that $\sigma_0^{\gamma(pn)}=2\sigma_0^{\gamma p}$. For the threshold behavior, $\sigma_{\gamma (pn)\to J/\psi}$ may be different from $\sigma_{\gamma p\to J/\psi}$, we will estimate the contributions for a wide range of $\beta_2=n_2\beta$ where $n_2$ in the range of $1-3$. 

\begin{figure}[h]
\centering
\includegraphics[width=0.95\columnwidth]{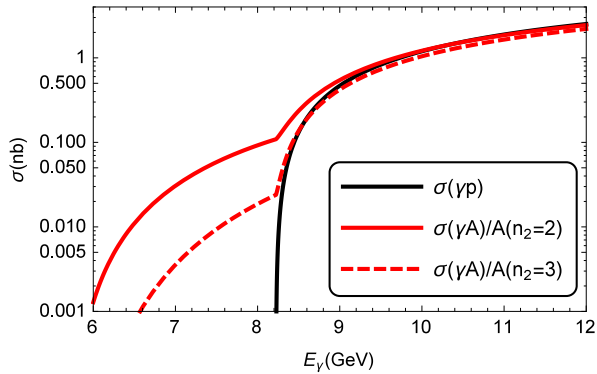}
\centering
\caption{Threshold and sub-threshold $J/\psi$ production in $\gamma p$ and $\gamma A$ collisions.}
\label{subthreshold}
\end{figure}

In Fig.~\ref{subthreshold}, we plot the near threshold and sub-threshold $J/\psi$ production cross sections in $\gamma p$ and $\gamma A$ collisions, respectively, based on the above formulas. For illustration purpose, we applied $n_2=2$ and $3$ for $\gamma A$ case. A number of corrections need to be taken care in the future. First, the nuclear absorption effects, which not only affect the above assumption, but also the individual terms in Eq.~(\ref{xspsi}). Second, the functional form may be totally different for the SRC contribution as compared to the $\gamma p$ cross section. Nevertheless, this provides a rough estimate, and we emphasize future measurement will definitely help us to understand the SRC cross section~\footnote{An early attempt has been made to measure $J/\psi$ production below the $\gamma p$-threshold with nuclear target at JLab~\cite{Bosted:2008mn}. However, due to too low energy ($E_\gamma\sim \rm 5.7 GeV$), there is no evidence of $J/\psi$ production. Hope future experiments at JLab can explore this in the sub-threshold region of $E_\gamma\sim 6-8\rm GeV$.}.

In addition, if we can accumulate enough events in the sub-threshold $J/\psi$ production, we may utilize the unique opportunity to explore the differential cross section respect to the momentum transfer $d\sigma/dt$. This will help to identify the SRC configuration, for example, the spatial size of the $(pn)$ pair in the nucleus. 

In Fig.~\ref{subthreshold}, we have neglected the nuclear absorption corrections for $J/\psi$ production cross sections in $\gamma A$ collisions. This can be completely avoided if we study open Charm production in the sub-threshold region. All of the above discussions can be extended to this process and other heavy flavor production processes, including $\Upsilon$s and open Bottom productions. Therefore, the universality of SRC implies that we have the following predictions,
\begin{eqnarray}
&&\frac{\tilde \sigma_{\gamma A\to J/\psi}}{\sigma_{\gamma d\to J/\psi}}\big|_{W_{\gamma p}<M_{pJ/\psi}}=\frac{\tilde\sigma_{\gamma A\to \Upsilon}}{\sigma_{\gamma d\to \Upsilon}}\big|_{W_{\gamma p}<M_{p\Upsilon}}\\
&&=    \frac{\sigma_{\gamma A\to c\bar c}}{\sigma_{\gamma d\to c\bar c}}\big|_{W_{\gamma p}<M_{pc\bar c}}=\frac{n_{src}^A}{n_{src}^d}=\frac{F_2^A(x_B)}{F_2^d(x_B)}|_{1.5<x_B<2.0} \ . \nonumber
\end{eqnarray}
For both $J/\psi$ and $\Upsilon$, we have included a nuclear absorption correction factor in the cross sections. This is a powerful prediction and can be tested in future experiments at JLab and IEEIC. In addition, the comparison between open Charm and Charmonium productions in the sub-threshold region shall provide useful information on the nuclear absorption effects for $J/\psi$. This information, in return, will help to reveal the nucleon-$J/\psi$ interaction and the trace anomaly contribution to the proton mass~\cite{Peskin:1979va,Bhanot:1979vb,Luke:1992tm,Kharzeev:1998bz,Kharzeev:1999jt,Brodsky:2000zc,Frankfurt:2002ka,Gryniuk:2016mpk,Hatta:2018ina,Hatta:2019lxo}. 

There has been an interesting proposal to study the SRC through exclusive $J/\psi$ production in electron-deuteron scattering process at the EIC~\cite{Miller:2015tjf}: $e+D\to e+J/\psi+p+n$, where a deuteron target disintegrates into a proton and a neutron with large relative momentum in the final state. The mechanism is very different from ours discussed above. The comparison between these two shall provide useful information on the underlying physics of the SRC in nucleus and their contributions to heavy flavor productions. 

\section{Conclusion}

In summary, we have studied the gluonic probe to the nucleon-nucleon short range correlation in nucleus at JLab and future intermediate energy EIC. Nuclear modification of the gluon distribution can be extensively investigated through the Charm structure function in the EMC region. Predictions based on EPPS16 and the universality argument have been presented. 

Furthermore, we have shown that the SRC can be explored through the subthreshold production of heavy flavor in $\gamma A$ collisions at JLab and intermediate energy EIC. In particular, the universality of the SRC predicts that the ratio of the nuclear cross sections can be directly linked to those measured in DIS structure functions.

{\it Acknowledgement.} We thank Yoshitaka Hatta for many discussions during the process of this project and a number of critical comments on the draft. We thank Jianping Ma and Nu Xu for discussions related to the EicC project. The material of this paper is based upon work partially supported by the LDRD program of Lawrence Berkeley National Laboratory, the U.S. Department of Energy, Office of Science, Office of Nuclear Physics, under contract number DE-AC02-05CH11231. J.X. is supported in part by the National Natural Science Foundation of China under Grant No. 11575110, 11735010, and 11911530088, the Natural Science Foundation of Shanghai under Grant No. 15DZ2272100, and the Key Laboratory for Particle Physics, Astrophysics and Cosmology, Ministry of Education of China.



\begin{thebibliography}{99}

\bibitem{Frankfurt:2008zv} 
  L.~Frankfurt, M.~Sargsian and M.~Strikman,
  Int.\ J.\ Mod.\ Phys.\ A {\bf 23}, 2991 (2008)
  doi:10.1142/S0217751X08041207
  [arXiv:0806.4412 [nucl-th]].
\bibitem{Arrington:2011xs} 
  J.~Arrington, D.~W.~Higinbotham, G.~Rosner and M.~Sargsian,
  Prog.\ Part.\ Nucl.\ Phys.\  {\bf 67}, 898 (2012)
  doi:10.1016/j.ppnp.2012.04.002
  [arXiv:1104.1196 [nucl-ex]].
  
\bibitem{Hen:2013oha} 
  O.~Hen, D.~W.~Higinbotham, G.~A.~Miller, E.~Piasetzky and L.~B.~Weinstein,
  Int.\ J.\ Mod.\ Phys.\ E {\bf 22}, 1330017 (2013)
  doi:10.1142/S0218301313300178
  [arXiv:1304.2813 [nucl-th]].
\bibitem{Hen:2016kwk} 
  O.~Hen, G.~A.~Miller, E.~Piasetzky and L.~B.~Weinstein,
  Rev.\ Mod.\ Phys.\  {\bf 89}, no. 4, 045002 (2017)
  doi:10.1103/RevModPhys.89.045002
  [arXiv:1611.09748 [nucl-ex]].
  
\bibitem{Fomin:2017ydn} 
  N.~Fomin, D.~Higinbotham, M.~Sargsian and P.~Solvignon,
  Ann.\ Rev.\ Nucl.\ Part.\ Sci.\  {\bf 67}, 129 (2017)
  doi:10.1146/annurev-nucl-102115-044939
  [arXiv:1708.08581 [nucl-th]].
  
\bibitem{Cloet:2019mql} 
  I.~C.~Cloet {\it et al.},
  arXiv:1902.10572 [nucl-ex].
  
\bibitem{Egiyan:2005hs} 
  K.~S.~Egiyan {\it et al.} [CLAS Collaboration],
  Phys.\ Rev.\ Lett.\  {\bf 96}, 082501 (2006)
  doi:10.1103/PhysRevLett.96.082501
  [nucl-ex/0508026].
  
\bibitem{Seely:2009gt} 
  J.~Seely {\it et al.},
  Phys.\ Rev.\ Lett.\  {\bf 103}, 202301 (2009)
  doi:10.1103/PhysRevLett.103.202301
  [arXiv:0904.4448 [nucl-ex]].
  
\bibitem{Weinstein:2010rt} 
  L.~B.~Weinstein, E.~Piasetzky, D.~W.~Higinbotham, J.~Gomez, O.~Hen and R.~Shneor,
  Phys.\ Rev.\ Lett.\  {\bf 106}, 052301 (2011)
  doi:10.1103/PhysRevLett.106.052301
  [arXiv:1009.5666 [hep-ph]].
\bibitem{Fomin:2011ng} 
  N.~Fomin {\it et al.},
  Phys.\ Rev.\ Lett.\  {\bf 108}, 092502 (2012)
  doi:10.1103/PhysRevLett.108.092502
  [arXiv:1107.3583 [nucl-ex]].
  
\bibitem{Hen:2012fm} 
  O.~Hen, E.~Piasetzky and L.~B.~Weinstein,
  Phys.\ Rev.\ C {\bf 85}, 047301 (2012)
  doi:10.1103/PhysRevC.85.047301
  [arXiv:1202.3452 [nucl-ex]].
  
\bibitem{Arrington:2012ax} 
  J.~Arrington, A.~Daniel, D.~Day, N.~Fomin, D.~Gaskell and P.~Solvignon,
  Phys.\ Rev.\ C {\bf 86}, 065204 (2012)
  doi:10.1103/PhysRevC.86.065204
  [arXiv:1206.6343 [nucl-ex]].
  
\bibitem{Hen:2014nza} 
  O.~Hen {\it et al.},
  Science {\bf 346}, 614 (2014)
  doi:10.1126/science.1256785
  [arXiv:1412.0138 [nucl-ex]].
  
\bibitem{Cohen:2018gzh} 
  E.~O.~Cohen {\it et al.} [CLAS Collaboration],
  Phys.\ Rev.\ Lett.\  {\bf 121}, no. 9, 092501 (2018)
  doi:10.1103/PhysRevLett.121.092501
  [arXiv:1805.01981 [nucl-ex]].
  
\bibitem{Duer:2018sby} 
  M.~Duer {\it et al.} [CLAS Collaboration],
  Nature {\bf 560}, no. 7720, 617 (2018).
  doi:10.1038/s41586-018-0400-z
  
\bibitem{Duer:2018sxh} 
  M.~Duer {\it et al.} [CLAS Collaboration],
  Phys.\ Rev.\ Lett.\  {\bf 122}, no. 17, 172502 (2019)
  doi:10.1103/PhysRevLett.122.172502
  [arXiv:1810.05343 [nucl-ex]].
  
\bibitem{Schmookler:2019nvf} 
  B.~Schmookler {\it et al.} [CLAS Collaboration],
  Nature {\bf 566}, no. 7744, 354 (2019).
  doi:10.1038/s41586-019-0925-9
  
\bibitem{Chen:2016bde} 
  J.~W.~Chen, W.~Detmold, J.~E.~Lynn and A.~Schwenk,
  Phys.\ Rev.\ Lett.\  {\bf 119}, no. 26, 262502 (2017)
  doi:10.1103/PhysRevLett.119.262502
  [arXiv:1607.03065 [hep-ph]].
  
\bibitem{Lynn:2019vwp} 
  J.~E.~Lynn, D.~Lonardoni, J.~Carlson, J.-W.~Chen, W.~Detmold, S.~Gandolfi and A.~Schwenk,
  arXiv:1903.12587 [nucl-th].
  
\bibitem{Lynn:2019rdt} 
  J.~E.~Lynn, I.~Tews, S.~Gandolfi and A.~Lovato,
  [arXiv:1901.04868 [nucl-th]].

\bibitem{Frankfurt:1993sp} 
  L.~L.~Frankfurt, M.~I.~Strikman, D.~B.~Day and M.~Sargsian,
  Phys.\ Rev.\ C {\bf 48}, 2451 (1993).
  doi:10.1103/PhysRevC.48.2451
\bibitem{Piasetzky:2006ai} 
  E.~Piasetzky, M.~Sargsian, L.~Frankfurt, M.~Strikman and J.~W.~Watson,
  Phys.\ Rev.\ Lett.\  {\bf 97}, 162504 (2006)
  doi:10.1103/PhysRevLett.97.162504
  [nucl-th/0604012].

\bibitem{Segarra:2019gbp} 
  E.~P.~Segarra, A.~Schmidt, D.~W.~Higinbotham, T.~Kutz, E.~Piasetzky, M.~Strikman, L.~B.~Weinstein and O.~Hen,
  arXiv:1908.02223 [nucl-th].
  
\bibitem{NuXu} Nu Xu, private communications.
  
\bibitem{Eskola:2016oht} 
  K.~J.~Eskola, P.~Paakkinen, H.~Paukkunen and C.~A.~Salgado,
  Eur.\ Phys.\ J.\ C {\bf 77}, no. 3, 163 (2017)
  doi:10.1140/epjc/s10052-017-4725-9
  [arXiv:1612.05741 [hep-ph]].
  
  
\bibitem{Kovarik:2015cma} 
  K.~Kovarik {\it et al.},
  Phys.\ Rev.\ D {\bf 93}, no. 8, 085037 (2016)
  doi:10.1103/PhysRevD.93.085037
  [arXiv:1509.00792 [hep-ph]].
  
\bibitem{Khanpour:2016pph} 
  H.~Khanpour and S.~Atashbar Tehrani,
  Phys.\ Rev.\ D {\bf 93}, no. 1, 014026 (2016)
  doi:10.1103/PhysRevD.93.014026
  [arXiv:1601.00939 [hep-ph]].
  
\bibitem{deFlorian:2011fp} 
  D.~de Florian, R.~Sassot, P.~Zurita and M.~Stratmann,
  Phys.\ Rev.\ D {\bf 85}, 074028 (2012)
  doi:10.1103/PhysRevD.85.074028
  [arXiv:1112.6324 [hep-ph]].
  

\bibitem{AbdulKhalek:2019mzd} 
  R.~Abdul Khalek {\it et al.} [NNPDF Collaboration],
  Eur.\ Phys.\ J.\ C {\bf 79}, no. 6, 471 (2019)
  doi:10.1140/epjc/s10052-019-6983-1
  [arXiv:1904.00018 [hep-ph]].
  
\bibitem{Walt:2019slu} 
  M.~Walt, I.~Helenius and W.~Vogelsang,
  arXiv:1908.03355 [hep-ph].
  
\bibitem{Aschenauer:2017oxs} 
  E.~C.~Aschenauer, S.~Fazio, M.~A.~C.~Lamont, H.~Paukkunen and P.~Zurita,
  Phys.\ Rev.\ D {\bf 96}, no. 11, 114005 (2017)
  doi:10.1103/PhysRevD.96.114005
  [arXiv:1708.05654 [nucl-ex]].
  
    
\bibitem{Arrington:2019wky} 
  J.~Arrington and N.~Fomin,
  arXiv:1903.12535 [nucl-ex].
  
\bibitem{Hen:2019jzn} 
  O.~Hen {\it et al.},
  arXiv:1905.02172 [nucl-ex].


\bibitem{Zhu:2007aa} 
  L.~Y.~Zhu {\it et al.} [NuSea Collaboration],
  Phys.\ Rev.\ Lett.\  {\bf 100}, 062301 (2008)
  doi:10.1103/PhysRevLett.100.062301
  [arXiv:0710.2344 [hep-ex]].
  
  
\bibitem{Peskin:1979va} 
  M.~E.~Peskin,
  Nucl.\ Phys.\ B {\bf 156}, 365 (1979).
  doi:10.1016/0550-3213(79)90199-8
  
\bibitem{Bhanot:1979vb} 
  G.~Bhanot and M.~E.~Peskin,
  Nucl.\ Phys.\ B {\bf 156}, 391 (1979).
  doi:10.1016/0550-3213(79)90200-1
  
\bibitem{Luke:1992tm} 
  M.~E.~Luke, A.~V.~Manohar and M.~J.~Savage,
  Phys.\ Lett.\ B {\bf 288}, 355 (1992)
  doi:10.1016/0370-2693(92)91114-O
  [hep-ph/9204219].
  
\bibitem{Kharzeev:1998bz} 
  D.~Kharzeev, H.~Satz, A.~Syamtomov and G.~Zinovjev,
  Eur.\ Phys.\ J.\ C {\bf 9}, 459 (1999)
  doi:10.1007/s100529900047
  [hep-ph/9901375].
\bibitem{Kharzeev:1999jt} 
  D.~Kharzeev, H.~Satz, A.~Syamtomov and G.~Zinovev,
  Nucl.\ Phys.\ A {\bf 661}, 568 (1999).
  doi:10.1016/S0375-9474(99)85090-8
  
\bibitem{Brodsky:2000zc} 
  S.~J.~Brodsky, E.~Chudakov, P.~Hoyer and J.~M.~Laget,
  Phys.\ Lett.\ B {\bf 498}, 23 (2001)
  doi:10.1016/S0370-2693(00)01373-3
  [hep-ph/0010343].
  
\bibitem{Frankfurt:2002ka} 
  L.~Frankfurt and M.~Strikman,
  Phys.\ Rev.\ D {\bf 66}, 031502 (2002)
  doi:10.1103/PhysRevD.66.031502
  [hep-ph/0205223].
  
\bibitem{Gryniuk:2016mpk} 
  O.~Gryniuk and M.~Vanderhaeghen,
  Phys.\ Rev.\ D {\bf 94}, no. 7, 074001 (2016)
  doi:10.1103/PhysRevD.94.074001
  [arXiv:1608.08205 [hep-ph]].
  
\bibitem{Hatta:2018ina} 
  Y.~Hatta and D.~L.~Yang,
  Phys.\ Rev.\ D {\bf 98}, no. 7, 074003 (2018)
  doi:10.1103/PhysRevD.98.074003
  [arXiv:1808.02163 [hep-ph]].
  
\bibitem{Hatta:2019lxo} 
  Y.~Hatta, A.~Rajan and D.~L.~Yang,
  arXiv:1906.00894 [hep-ph].
  
\bibitem{Ji:1994av} 
  X.~D.~Ji,
  Phys.\ Rev.\ Lett.\  {\bf 74}, 1071 (1995)
  doi:10.1103/PhysRevLett.74.1071
  [hep-ph/9410274].
    
\bibitem{Anderson:1976hi} 
  R.~L.~Anderson {\it et al.},
  Phys.\ Rev.\ Lett.\  {\bf 38}, 263 (1977).
  doi:10.1103/PhysRevLett.38.263


\bibitem{Gittelman:1975ix} 
  B.~Gittelman, K.~M.~Hanson, D.~Larson, E.~Loh, A.~Silverman and G.~Theodosiou,
  Phys.\ Rev.\ Lett.\  {\bf 35}, 1616 (1975).
  doi:10.1103/PhysRevLett.35.1616
  
\bibitem{Camerini:1975cy} 
  U.~Camerini {\it et al.},
  Phys.\ Rev.\ Lett.\  {\bf 35}, 483 (1975).
  doi:10.1103/PhysRevLett.35.483

\bibitem{Ali:2019lzf} 
  A.~Ali {\it et al.} [GlueX Collaboration],
  arXiv:1905.10811 [nucl-ex].
  


\bibitem{Bosted:2008mn} 
  P.~Bosted {\it et al.},
  Phys.\ Rev.\ C {\bf 79}, 015209 (2009)
  doi:10.1103/PhysRevC.79.015209
  [arXiv:0809.2284 [nucl-ex]].
  
\bibitem{Miller:2015tjf} 
  G.~A.~Miller, M.~D.~Sievert and R.~Venugopalan,
  Phys.\ Rev.\ C {\bf 93}, no. 4, 045202 (2016)
  doi:10.1103/PhysRevC.93.045202
  [arXiv:1512.03111 [nucl-th]].

\end{thebibliography}
\end{document}